\documentclass[%
 reprint,
 amsmath,amssymb,
 aps,
]{revtex4-2}
\usepackage{graphicx}
\usepackage{xcolor}
\usepackage{dcolumn}
\usepackage{bm}
\usepackage[nopatch=footnote]{microtype} 
\usepackage{float}
\usepackage{placeins}
\usepackage{gensymb}
\usepackage{hyperref}
\usepackage[T1]{fontenc} 

\begin{document}

\preprint{APS/123-QED}

\title{Impact of particle-size polydispersity on the quality of thin-film colloidal crystals}

\author{Mariam Arif}
 \email{mariam.arif@ed.ac.uk}
\affiliation{%
School of Physics and Astronomy, The University of Edinburgh,  Peter Guthrie Tait Road, Edinburgh, EH9 3FD, Scotland, United Kingdom
}%
\author{Andrew B. Schofield}
\affiliation{%
School of Physics and Astronomy, The University of Edinburgh,  Peter Guthrie Tait Road, Edinburgh, EH9 3FD, Scotland, United Kingdom
}%
\author{Fraser H. J. Laidlaw}
\affiliation{%
 School of Physics and Astronomy, The University of Edinburgh,  Peter Guthrie Tait Road, Edinburgh, EH9 3FD, Scotland, United Kingdom
}%
\author{Wilson C. K. Poon}
\affiliation{%
 School of Physics and Astronomy, The University of Edinburgh,  Peter Guthrie Tait Road, Edinburgh, EH9 3FD, Scotland, United Kingdom
}%
\author{Job H. J. Thijssen}%
 \email{j.h.j.thijssen@ed.ac.uk}
\affiliation{%
 School of Physics and Astronomy, The University of Edinburgh,  Peter Guthrie Tait Road, Edinburgh, EH9 3FD, Scotland, United Kingdom
}%

\date{\today}

\begin{abstract}
Size polydispersity in colloidal particles can disrupt order in their self-assembly, ultimately leading to a complete suppression of crystallization. In contrast to various computational studies, few experimental studies systematically address the effects of size polydispersity on the quality of colloidal crystals. We present an experimental study of structural order in thin films of crystals vertically dried from colloidal dispersions with a systematically varying polydispersity. As expected, an increase in polydispersity leads to a deterioration in order with significant drops in the local bond-orientational order at 8\% and 12\% polydispersity. Our results align with previously suggested models of epitaxial-like growth of 2D layers during convective assembly. Our results can offer critical insights into the permissible limits for achieving colloidal crystals from more polydisperse systems such as those synthesized through more sustainable methods.
\end{abstract}

\maketitle


\section{\label{sec:Intro}Introduction}
Colloidal crystals have attracted significant interest due to their ease of preparation and range of applications. Conventionally, they have been used as photonic nanocrystals \cite{kim2011self}, for which the photonic bandgaps of these materials are crucial \cite{rengarajan2005effect} and so necessitating high crystallinity \cite{ye2001self, li2019recent}. Other applications, which do not utilize the optical properties of these crystals, such as their use as templates for microbatteries \cite{sorensen2006three, pikul2013high}, may allow greater flexibility in crystal preparation and particle synthesis techniques as the requirements for crystal quality might be less demanding.
\par To optimize colloidal crystal quality, model systems of particles with low polydispersity are often employed. One of the most popular colloid synthesis techniques is the Stöber synthesis, as it provides good control of particle size  \cite{bourebrab2018influence, fernandes2019revising} and the resulting batches can have polydispersity as low as 4-5\% \cite{giesche1994synthesis}. In this method, a silica precursor undergoes hydrolysis in a water-alcohol mixture catalysed by ammonia which is followed by condensation of silanol groups to form siloxane networks. With recent attempts to make the Stöber synthesis sustainable by using water as the primary solvent \cite{wang2023modifying, furlan2019water}, it is a challenge to control the reaction and uniformity of the resulting particles \cite{wang2023modifying}. The non-uniformity in particle size disrupts regularity in the arrangement of colloidal crystals \cite{topccu2018non}, which can be a significant impediment to maintaining their quality. This deterioration in crystal quality, however, is yet to be studied quantitatively over a range of particle size polydispersities to better identify and understand critical thresholds. Such a study can also give more insight into the assembly processes. 
\par Previous experimental work on the drop in crystal quality with increasing polydispersity has focused on relatively low polydispersities \cite{rengarajan2005effect}  and on the variation of optical behaviour of crystals \cite{harun2010angle}. The long-range structural order, up to an upper limit of 10\% polydispersity, was investigated by Rengaranjan \textit{et al.} in \cite{rengarajan2005effect}. In another study, this variation was characterised in Fourier space for two silica batches, with  narrow and wide particle size distributions \cite{jiang1999single}. The limited scope of systematic previous experimental work could stem from the challenging nature of synthesizing a finely spaced range of polydispersities without significantly altering the average particle diameter. Such a challenge need not be met in non-experimental studies, so that the impact of polydispersity on crystal quality for hard sphere colloids has been studied theoretically \cite{pusey1987effect} and through computer simulations \cite{auer2001suppression, phan1998effects}, with crystallization suppression reported at polydispersity levels between 6-12\%. To the best of our knowledge, there has been no accurate determination of this limit using experiments, but only reports that crystallization is not observed at particular polydispersities. The variation of local order as polydispersity is increased towards this limit also does not appear to have been systematically investigated in experiments.
\par Here we systematically investigate structural order variation in thin films of colloidal crystals fabricated by vertical drying using silica particles with polydispersities ranging from 6\% to 15\%; this range aligns with previous reports of significant shifts in ordering behaviour \cite{pusey1987effect, auer2001suppression, van2013effect}. The colloidal dispersions are characterised using electron microscopy and light scattering, while the resulting (crystalline) assemblies are characterised in 2D and 3D using electron microscopy and UV-Vis spectroscopy. As expected, both long-range and local order decrease with increasing polydispersity. Notably, we observe two significant drops in local order around 8\% and 12\% polydispersity. In addition, we observe that the local order is higher in crystal layers further from the substrate. We also measure the polydispersity of particles in the (crystalline) assembly, and in the dispersion before and after crystal growth, and do not find significant size-fractionation. We discuss these results in the context of existing literature, arguing that they can be explained by known polydispersity limits for 2D and 3D crystallization \cite{kawasaki2008link, pusey1987effect}, operating in the context of a model for convective assembly of epitaxial-like growth of 2D crystalline layers to form 3D thin-film colloidal crystals \cite{wang2025situ}.
\section{\label{sec:Methods}Experimental Methods}
\subsection{Synthesis of silica particles}
Silica nanoparticles were synthesized following the Stöber-Fink-Bohn method \cite{stober1968controlled}. Ethanol (Sigma Aldrich, $\geq$99.8\%), distilled water, tetraethyl orthosilicate (TEOS, Sigma Aldrich, $\geq$99\%) and ammonia solution (Fisher Chemical, 35\% in H$_2$O) were mixed and left for 48 hours to react at room temperature and pressure. A systematic variation of reactant volumes, as discussed in references \cite{fernandes2019revising} and \cite{ghimire2021renaissance}, enabled the production of low (6.3-7.5\%) and high (11.9\%, 14.6\%) polydispersity silica batches within a consistent particle diameter range of 200-400 nm. Two `intermediate' polydispersity batches (8.5\% and 10.8\%) were made by mixing (almost) equal number concentrations of batches for which the average diameters lay within 2 standard deviations of each other. The 9.1\% polydisperse silica were made following the low polydispersity silica synthesis recipe but by a batch-wise addtion of TEOS over a period of 5 hours. All dispersions were cleaned by performing 7 successive cycles of centrifugation (using Thermo Scientific Sorvall ST 40), supernatant removal and ethanol addition. The recipes for all batches can be found in the Supplementary Information.
\subsection{Characterization of particles}
The average hydrodynamic radii of our silica particles were measured by Anton Paar Litesizer DLS 500 using Kalliope software. The particle size distributions were determined by Scanning Electron Microscopy (SEM) using JEOL JSM-6010PLUS/LV. An SEM sample was prepared by drying a droplet of dispersion (10-12\% volume in ethanol) on a silicon wafer substrate stuck to a metal stub with carbon tape. The top surface of the dried sample was sputter coated with a 20 nm layer of gold using a Denton Vacuum Desk III sputter coater. The surface was imaged using the SEM (Figure~ \ref{fig:Silica}(a), (b)) at different locations across the sample and particle sizes for 1000 particles were manually measured in ImageJ (version 1.54f) using 6-10 images from each batch. The size distribution histogram is plotted (Figure~ \ref{fig:Silica}(c), (d)) and the polydispersity (PD) of the batch is obtained by taking the ratio of standard deviation ($\sigma_d$) of the distribution of diameters to the mean particle diameter ($d_{\text{avg}}$), expressed as a percentage:
\begin{equation} \label{pd}
\mathrm{PD} = \frac{\sigma_d}{d_{\text{avg}}} \times 100\%
\end{equation}
An overview of SEM and DLS results for our particle batches is presented in Table~ \ref{tab:silica_table}. The polydispersity from SEM was also verified, for a few particle batches, by performing Static Light Scattering (SLS) using the ALV/ CGS-3 Compact Goniometer. A He-Ne beam (632.8 nm) was used as the light source and the detector was rotated from 30\,\textdegree\ to\ 150\,\textdegree\ in\ intervals\ of\ 1\,\textdegree. Experimental data is fitted to a theoretical  Mie curve, generated using Mieplot \cite{Laven_2012}, to obtain values for average particle size and polydispersity.
\par The average particle densities for two silica batches were also measured; we chose two batches that were used to create a mixed `intermediate' batch. For each of the two batches, five silica alcosols were prepared, ranging from 0.3\% to 3\% mass fraction, and the densities of these were measured using a density meter (Anton Paar DMA 4500). Assuming conservation of volume, and applying the density-mass-volume relationship, the following expression for dispersion density ($\rho_t$) in terms of mass fraction ($\phi_m$) can be derived:
\begin{equation} \label{densityeq}
\frac{1}{\rho_t} = \phi_{\text{m}} \left(\frac{1}{\rho_p}-\frac{1}{\rho_s}\right) + \frac{1}{\rho_s} 
\end{equation}
where $\rho_p$ is the particle density and $\rho_s$ is the solvent density. Hence, the solvent density can be obtained from the intercept and the particle density via the slope of a ($\frac{1}{\rho_t}$, $\phi_{m}$) plot. The solvent density can also be measured independently and compared to $\rho_s$ to validate this procedure for measuring particle density.
\subsection{Fabrication of colloidal crystals}
Crystals were fabricated by vertical drying following Jiang \textit{et al.} \cite{jiang1999single} as the method is robust and reliable. Glass microscopy slides (Thermo scientific; 75×25×1 mm) cut up lengthwise were used as substrates for crystal growth. These were cleaned with ethanol and dried prior to use. Silica alcosols of 0.5-1\% volume fraction were prepared and 7 ml of the dispersion was added to a 28 ml glass vial (Samco, trident screw neck vial). Slides were carefully placed in these vials such that they were vertical (see  Figure~ \ref{fig:crystals}(a)). The setup was covered and left in a fume hood for seven days before retrieving the samples. Temperature and humidity were regularly monitored during the crystal growth process. The average growth temperature and relative humidity were (19.7 ± 0.5)\degree C and (39 ± 4)\% respectively for the samples used in this study.
\subsection{Characterization of assemblies}
The retrieved slides were cleaned on one side with ethanol and sputter coated on the other side with a 15-20 nm gold layer to image the assemblies at different locations in the sample using the JEOL SEM. For characterization in 2D, these images were processed in ImageJ to manually mark the position of particle centroids for 400-500 particles per image for 7 images per batch. Multiple samples were grown from each batch and the reproducibility of crystal quality was evaluated by analysing images from three samples for some of the batches.
\par To assess long-range positional correlation for these assemblies, the radial distribution function was calculated. It has been used as a common measure to quantify order in colloidal assemblies \cite{payam2015quantitative} and is expressed in 2D as:
\begin{equation} \label{RDFequation}
g(r) =  \frac{n(r)}{ 2 \pi r \rho_n \delta r }
\end{equation}
where \textit{n(r)} is the number of particles in an annular region of inner radius \textit{r}, $\delta r$ is the width of this region and $\rho_n $ is the particle number density. To implement this, the particle centroids were used and the number of particles that lie within an annular region of width $\delta r$ at a distance \textit{r} away from an origin particle were counted. The annular width ($\delta r$) was determined by the pixel size of the image which depends upon the magnification of each image. This calculation was repeated considering each particle within the central 70\% of the image as origin particle in turn. This limit was applied to mitigate any finite image size effects. These counts were normalized by $\rho_n $ and by the area of the respective annular regions, as shown in equation (\ref{RDFequation}). The result was divided by the $g(r)$ of an ideal gas, calculated using the same number of particles but with randomized positions. The resulting $g(r)$ is plotted as a function of the radial distance normalized by half of the nearest neighbour distance for each batch so nearest neighbours are, on average, at a normalized distance of 2. 
\par To analyze short-range variation in order, the local bond-order parameter was used. For a single particle positioned at $x_j$, this can be calculated as \cite{gribova2011close}:
\begin{equation} \label{sixBOP}
\psi_s(x_j) = \left| \frac{1}{N_j}\sum_{k=1}^{N_j} \exp(si\theta_{jk})\right|
\end{equation}
where s is the fold-symmetry, $N_j$ is the number of nearest neighbours and $\theta_{jk}$ is the angle between a line connecting the centroids of the nearest neighbour (\textit{k}) and the reference particle (\textit{j}) and a reference axis. For perfect 6-fold symmetry, $\psi_6$  is unity and decreases as disorder increases. The local six-fold bond-order parameter for an ensemble of particles ($\Psi_6$) is the mean of $\psi_6$ values for all the particles in the system. For assemblies made from each of the silica batches listed in Table~ \ref{tab:silica_table}, $\Psi_6$ was calculated for 7 randomly selected SEM images and averaged ($\left<\Psi_6\right>$). The nearest neighbours for each particle in an image were determined by Delaunay triangulation. As the particles at the boundary of the region in consideration do not appear to have neighbours, this was identified as an edge effect and is addressed by excluding boundary particles from being considered as origin particles.
\par To view stacking in the assemblies, the samples were cut using Focused Ion Beam (FIB) milling in a Zeiss Crossbeam 550 FIB-SEM. A 20×3×1-2 $\mu$m platinum coating was deposited in the centre of the sample and a 6-7 $\mu$m deep trench was milled at this location to expose the cross section. The degree to which assemblies have planes of particles with well-defined interplanar spacing was characterised by measuring their transmission spectra using a UV-Vis Spectrophotometer (Agilent Cary 100 double beam UV-Vis) and Scan software. A black tape with 1 mm hole punched through it covered the empty side of the glass slide, ensuring that the beam strictly passes through the desired area in the centre of the sample to measure its transmitted intensity in a wavelength range of 300-900 nm. The intensities were normalized by the incident intensity on the sample, defined as the transmitted intensity through only the glass slide via the 1 mm hole. 
\par The 3D assembly was further investigated through a layer-by-layer analysis of the average local six-fold bond-order parameter ($\left<\Psi_6\right>$) of the low polydispersity (6.3\%) assembly. The first, second and top-most layers in the assembly were imaged and the local six-fold bond-order parameter was calculated for 400-500 particles in each image. The results were then averaged across five images for each layer. Subsequent layers were accessible for imaging near the upper edge of the sample, where crystal growth begins during convective assembly. In this region, the colloidal assembly appeared step-like, clearly separated by square-patterned transitional zones, which are features characteristic of colloidal crystals grown via convective assembly\cite{hu2021structural}.
\section{\label{sec:R and D}Results and Discussion}
\par
\begin{figure}
\includegraphics{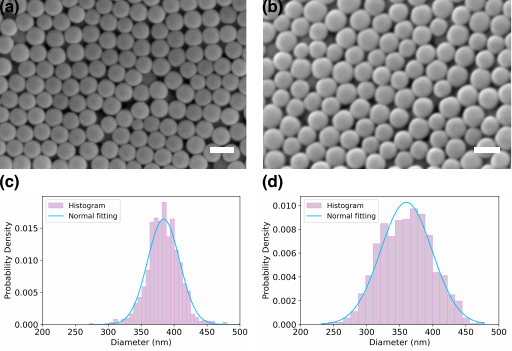}
\caption{\label{fig:Silica} SEM images of (a) 6.5\% and (b) 10.8\% polydispersity silica. The scale bars are 0.5 $\mu$m. (c), (d) The corresponding particle size distributions, from measuring 1000 particle diameters, fitted with normal distribution.}
\end{figure}

 The synthesized silica particles are typically spherical as seen in the SEM micrographs in Figure~ \ref{fig:Silica}(a),(b). Two of the particle size distributions calculated from such images, corresponding to the 6.5\% and 10.8\% polydispersity silica batches, fitted with normal distributions are presented in Figure~ \ref{fig:Silica}(c),(d). All the distributions are monomodal and resemble a Gaussian curve. This is especially important to check for mixtures, such as the 10.8\% polydispersity silica (Figure~ \ref{fig:Silica}(b),(d)). As the diameters of the mixed batches lie close to one another, mixing in equal concentrations enables the formation of a resultant distribution with a single peak. The particle densities of the two individual batches that were used to make the 10.8\% polydispersity mixture are also measured. These are 1.90 g/ml and 2.03 ± 0.04 g/ml respectively, which are also consistent with the literature values for Stöber silica's density \cite{badley1990surface, bell2012emerging}. This renders them favorable for mixing and suggests no density-based segregation of the two batches, as is also seen in the representative SEM image of the mixture (Figure~ \ref{fig:Silica}(b)).
\begin{table}
\centering
\begin{tabular}{|c|c|c|c|c|}
\hline
Batch & PD (\%) & SEM size (nm) & DLS size (nm) & \% difference \\ \hline
1 & 6.3 & 384 & 432 & 12.5 \\ \hline
2 & 6.5 & 322 & 377 & 17.1 \\ \hline
3 & 7.5 & 381 & 420 & 10.2 \\ \hline
4 & 8.5 & 371 & 422 & 13.7 \\ \hline
5 & 9.1 & 365 & 412 &  12.9 \\ \hline
6 & 10.8 & 360 & 410 &  13.9\\ \hline
7 & 11.9 & 208 & 260 & 25.0 \\ \hline
8 & 14.6 & 208 & 266 & 27.9 \\ \hline
\end{tabular}
\caption{Polydispersity and average particle diameter results for the eight silica batches included in this study, as determined from particle size distributions using Scanning Electron Microscopy (SEM) and from Dynamic Light Scattering (DLS). The last column gives the percentage difference between diameters measured from the two techniques.}
\label{tab:silica_table}
\end{table}
\begin{figure*}[!htbp]
\centering
\includegraphics[width=\linewidth]{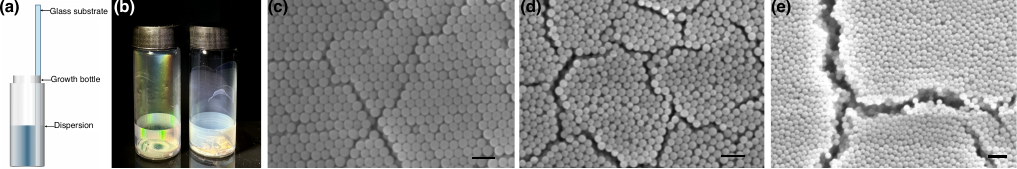} 
\caption{\label{fig:crystals} (a) Schematic of vertical drying setup. (b) Photograph of vials containing low (6.3\%, left) and high (14.6\%, right) polydispersity colloidal assemblies on the interior, showing difference in appearance (iridescence) of samples. SEM images (top view) of self-assemblies of (c) 6.3\%, (d) 9.1\% and (e) 14.6\% polydispersity silica batches. All scale bars are 1 $\mu$m.}
\end{figure*}
\par Table~ \ref{tab:silica_table} shows the polydispersities and average particle diameters of all the prepared silica batches. The difference between DLS and SEM sizes is a consequence of the DLS measuring hydrodynamic particle radius and additional shrinkage due to vacuum in SEM. DLS diameters for Stöber silica have therefore been reported to be around 20\% higher than those obtained from SEM \cite{topccu2018non}. The SEM sizes are also manually remeasured for 1000 particles from the same batch to investigate the error in manual measurement. The average particle diameters vary by ±5 nm and the discrepancy in polydispersity is found to be ±0.3\%-points. To ensure that the SEM characterization performed on particles in vacuum is representative of their state during crystal fabrication (in dispersion), the polydispersity and average particle size of silica in solution for some batches are also measured by static light scattering. These agree with the results in Table~ \ref{tab:silica_table}, for example for batch 3, the SLS diameter and polydispersity are 426 nm and 8\%.

\par Figure~\ref{fig:crystals}(a) shows a schematic of the crystallization setup and \ref{fig:crystals}(b) presents a side-by-side comparison of a low and high polydispersity growth bottle after the glass slides have been removed. The dispersion dried on the inner wall of these bottles shows the first sign in their difference in structural order. The appearance of structural colors at specific angles of observation (iridescence) due to constructive interference from crystal layers can be seen in the low polydispersity growth bottle (left) but is absent in the high polydispersity sample (right). The SEM micrographs in Figure~\ref{fig:crystals} show representative images of assemblies of (c) low (6.3\%), (d) intermediate (9.1\%) and (e) high (14.6\%) polydispersity silica. The low polydispersity particles are typically arranged uniformly, forming hexagonally packed structures as seen in Figure~\ref{fig:crystals}(c). As the polydispersity is increased to the intermediate range, the assemblies visibly start to lose hexagonal order. At even higher polydispersities, the 2D order seems to vanish and the particle positions appear more random, as seen in Figure~\ref{fig:crystals}(e).
\par
\begin{figure*}[!htbp]
\centering
\includegraphics[width=\linewidth]{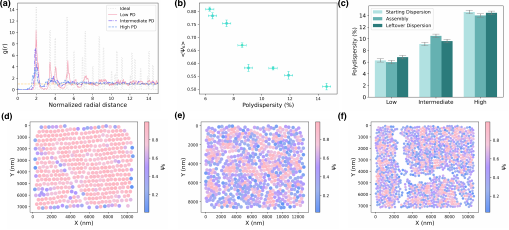}
\caption{\label{fig:RDFandBOP} Characterization of 2D order in assemblies. (a) Radial distribution function ($g(r)$) computed from SEM images of assemblies of low (6.3\%), intermediate (9.1\%) and high (14.6\%) polydispersity silica particles. The grey lines show the peak positions for a perfectly hexagonal system with the same (average) number of particles. The dashed line at $g(r)$=1 shows the convergence of the functions at larger radial distances. The horizontal axis has been scaled such that the position of the first peak is 2. (b) Average local six-fold bond-order parameter ($\left< \Psi_6 \right>$) for assemblies made from silica with polydispersities ranging from 6-15\%. (c) A comparison of the polydispersity (in dried samples) from the starting dispersion, assembly and leftover dispersion (after vertical drying) for low, intermediate and high polydispersity silica batches. (d-f) $\psi_6$ of particles in 6.3\%, 9.1\% and 14.6\% polydispersity silica assemblies computed from SEM images in Figure~ \ref{fig:crystals}(c-e).}
\end{figure*}

 To observe the shift in structural long-range order, we calculate the radial distribution function from particle centroids as described in section \ref{sec:Methods}. Figure~ \ref{fig:RDFandBOP}(a) presents the results for assemblies from three batches with low (6.3\%), intermediate (9.1\%) and high (14.6\%) polydispersity. The grey lines represent the $g(r)$ for an ideal hexagonal close packed lattice, where peaks appear periodically as a series of $\delta$ functions; the width of these peaks is set to a minimal width here for visual clarity. The $g(r)$ for low polydispersity silica assembly shows the most distinct peaks out of the three experimental data sets, indicating the presence of the highest degree of long-range order. For the intermediate polydispersity assembly, $g(r)$ shows fewer, less distinct peaks of lower amplitude than the low polydispersity one, suggesting a loss of long-range order. For the high polydispersity assembly, there is an absence of any prominent peaks after the first, suggesting the absence of any substantial crystalline order. Instead, this $g(r)$ closely resembles that of a shear melted colloidal crystal \cite{bevan2004structural} or other low order systems such as particles in a gas-solid fluidized bed \cite{wang2009prediction}. For the low (6.3\%) polydispersity assembly (pink solid line in Figure~ 3(a)), it is also noted that the peaks in $g(r)$ at normalized radial distances beyond 4.0 do not always align with the $g(r)$ for an ideal hexagonal lattice but this has been observed previously for colloidal crystals grown by vertical drying \cite{rengarajan2005effect} and may be related to the previously reported observation that the hexagonal layers in vertically dried colloidal crystals are expanded along the meniscus direction by 4\% \cite{Thijssen2007thesis}.
\par To quantify short-range order in 2D, we present the average local six-fold bond-order parameter ($\left< \Psi_6 \right>$) in Figure~ \ref{fig:RDFandBOP}(b), for assemblies from all batches listed in Table~ \ref{tab:silica_table}. The low polydispersity silica particles produce crystals in which $\left< \Psi_6 \right>$ is relatively close to that of a perfectly hexagonal lattice. As the polydispersity increases, $\left< \Psi_6 \right>$ decreases slowly to a point of inflection at around 7\%  followed by a sharp decline to a plateau at $\approx 0.58$ between 9-12\%, before dropping again to 0.50 at 14.6\%. Figure~ \ref{fig:RDFandBOP}(d-f) show $\psi_6$ for individual particles in the SEM images from Figure~ \ref{fig:crystals}(c-e), respectively.
\par Assemblies from our low polydispersity batch have a $\left< \Psi_6 \right>$ that is $\lesssim 20\%$ lower than that of a perfect hexagonal lattice. This is not unexpected given that the polydispersity of this batch is just above the upper bound (6\%) found by Rengaranjan \textit{et al.} \cite{rengarajan2005effect} for growing ``high quality'' silica colloidal crystals. However, since an order parameter of 0.7 has been quoted as a valid threshold for 6-fold order \cite{payam2015quantitative} our results for low-polydispersity samples can be considered as being within the crystalline regime.
\par To ensure that the trend seen in Figure~ \ref{fig:RDFandBOP}(b) is truly from the variation in polydispersity and not from the difference in average diameters of silica batches, $\left< \Psi_6 \right> $ is also plotted against average particle diameter (see Supplementary Information). This shows a random variation and no noticeable trend, confirming that for our system, the particle size does not drive the variation in structural order. The degree of order is found to be highly reproducible for multiple crystals made from the same silica batch. For example, the sample-to-sample variability in $\left< \Psi_6 \right>$ for 3 assembly samples of the 6.3\% batch is found to be ±0.009 and for 14.6\% crystals, this is ±0.008.
\par The shallow but clearly noticeable drop in the average local six-fold bond-order parameter $\left< \Psi_6 \right>$ at around 12\% polydispersity in Figure~\ref{fig:RDFandBOP}(b) can be explained using an argument analogous to the Lindemann criterion \cite{pusey1987effect}: crystal formation is impeded when the average particle radius plus the corresponding standard deviation is larger than half the lattice spacing. Starting from the melting volume fraction for hard spheres ($\phi = 0.545$), the polydispersity above which crystallization is impeded can then be calculated as approximately 11\%. Note that our system consists of charged rather than hard-sphere particles, but charging effects are expected to be relatively small (see below), so that it is not unreasonable to compare their behaviour to hard spheres.
\par The more significant drop in average local six-fold bond-order parameter $\left< \Psi_6 \right>$ between 7.5\% and 9.1\% polydispersity aligns with previous reports that dispersions of hard spheres with a polydispersity beyond 7.0\% to 8.5\% fractionate in order to crystallize \cite{Botet2016FD, Palberg2014JPCM, Fairhurst1999thesis, Bartlett1998Frac, Pusey2009Phil, pusey1991liquids}. To investigate fractionation in our system, we compare the polydispersity of low (6.3\%), intermediate (9.1\%) and high (14.6\%) polydispersity batches as measured from SEM images of the (dried) dispersion itself, of the resulting (crystal) assemblies, and of the (dried) dispersion left behind in the vial after removing the slide (see Figure~\ref{fig:RDFandBOP}(c)). These results confirm that for our chosen size range there was little size segregation of particles during the vertical drying. This can perhaps be explained by the vertical drying process in which the evaporation leads to a convective flow towards the meniscus region and the particles are swept along. During this process, the particles have little time to fractionate, so that crystallization will be strongly hindered for dispersions with a polydispersity larger than 7.0\% to 8.5\%, for which fractionation is needed for normal crystallization at equilibrium \cite{Botet2016FD, Palberg2014JPCM, Fairhurst1999thesis, pusey1991liquids, Pusey2009Phil, Bartlett1998Frac}.
\par An alternative explanation for the drop in average local six-fold bond-order parameter between 7.5\% and 9.1\% relies on previous reports that convective-assembly processes involve a 2D crystallization stage \cite{shimmin2006slow, wang2025situ}. For example, Wang \textit{et al}.~reported that the assembly process during dip coating, which is also a form of convective assembly, involves the formation of 2D monolayers at the solvent-air interface followed by epitaxial deposition onto the substrate forming a 3D crystal \cite{wang2025situ}. For hard disks in 2D, melting and/or crystallization suppression limits have previously been reported to be around 8\% polydispersity \cite{pronk2004melting, kawasaki2008link}, coinciding with the position of the sharp drop observed in Figure~ \ref{fig:RDFandBOP}(b).
\par
\begin{figure}
\includegraphics{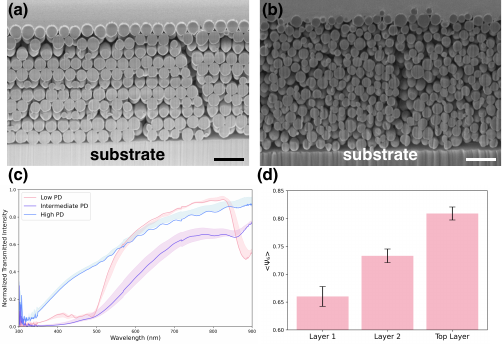}
\caption{\label{fig:TS} Characterization of 3D order in assemblies. (a), (b) SEM view of cross section of assemblies from low (6.3\%) and high (14.6\%) polydispersity silica with position of the substrate labeled. The scale bars are 1 $\mu$m. The faint vertical lines seen in (b) are an imaging artefact (`curtaining'), which is typical in FIB-SEM when encountering increased surface roughness \cite{giannuzzi2004introduction}. (c) Transmission spectra of assemblies of low (6.3\%), intermediate (9.1\%) and high (14.6\%) polydispersity silica. The transmission intensity has been normalized by incident intensity. The errors, shown as shaded regions, have been calculated from multiple runs, with each run performed on a different sample fabricated from the same particle batch. (d) Average local six-fold bond-order parameter ($\left <\Psi_6 \right>$) for layers in a low polydispersity (6.3\%) assembly sample.}
\end{figure}

 To distinguish between these two explanations, we consider the 3D structure of our samples. Figure~ \ref{fig:TS}(a),(b) show SEM images of a cross section through the crystal displaying how the particles are stacked in low (6.3\%) and high (14.6\%) polydispersity assemblies. The loss of order previously found in 2D is also apparent in 3D. The low polydispersity crystal shows well defined, regularly stacked layers whereas the high polydispersity assembly shows no particular stacking pattern and wider gaps owing to the size disparity of stacked particles.
\par The transmission spectra in Figure~ \ref{fig:TS}(c) show how these layers interact with light. For the low polydispersity assembly, back scattering from the periodic planes oriented perpendicular to the light beam result in a relatively sharp dip corresponding to a first order Bragg peak. The wavelength ($\lambda_{\text{peak}}$) of this peak at 883 nm corresponds to an interlayer spacing of 332 nm (considering the plane to be (111) in face-centred cubic indexing) for particles of average diameter of $\approx 407$~nm at close packing. This is consistent with the size of batch 1 particles (reported in Table~ \ref{tab:silica_table}) given that the assemblies are dried in air so the particle diameter is expected to lie in between the diameter in solution (DLS) and the diameter in vacuum (SEM). For the intermediate polydispersity silica assembly, the first order Bragg peak is broadened and much less pronounced than that for low polydispersity assembly, with a $\lambda_{\text{peak}}$ of 856 nm. The broadened peak indicates a significant loss of order in 3D which results in much weaker constructive interference. This loss of periodicity in 3D was also reported \cite{rengarajan2005effect} to occur at about 7\% polydispersity through an analysis of the reflectivity spectra of assemblies with varying polydispersities. For the 14.6\% polydispersity assembly, only the Fabry-Perot fringes can be identified which are formed from the light reflected back and forth between the substrate-assembly and air-assembly interfaces. This indicates ill-defined interlayer spacing, which is expected from irregularly stacked layers.
\par Figure. \ref{fig:TS}(d) shows the average local six-fold bond-order parameter for layers in a low polydispersity assembly, revealing an increase in local order from the bottom to the top. The layers are clearly distinguishable as seen in Fig 4(a) and have a strong crystalline order in 3D, as suggested by their clear Bragg peak in the transmission spectrum in Fig \ref{fig:TS}(c). We hypothesize that the top-down loss in local order, seen in Figure~ \ref{fig:TS}(d) may have arisen because the first layer (`Layer 1') is deposited directly on and adhered to the rigid substrate, which therefore preserves the order it had at the time of its formation. The degree of ordering in this layer will therefore be `quenched' from the beginning. By contrast, we expect the second layer deposited onto the first layer could rearrange by Brownian motion for a period of time after deposition, allowing a certain amount of annealing and therefore improving its degree of order. The increase in
bond-orientational order for layers further away from the
substrate indicates epitaxial-like
growth of 2D layers, previously suggested for convective assembly \cite{shimmin2006slow, wang2025situ}.
\par As our system consists of charged rather than hard-sphere particles, we should consider the role of charge. To do so, we first measured the zeta potential and found it to be consistent across batches, with an average value of (-49.0 $\pm$ 1.1) mV (see Supplementary Information). Second, we estimate the Debye length of the particles by estimating the crystalline fraction within a bulk sediment. This was done by centrifuging down a 10\% volume fraction dispersion of the lowest polydispersity (6.3\%) sample. After leaving it undisturbed for a few days, Bragg colours appeared throughout more than 89\% of the sediment. The total number of silica particles in the sample before sedimentation is equal to the number of silica particles in the crystalline fraction (effective packing fraction = 0.545 \cite{Pusey2009Phil}) plus those in the glassy fraction (effective packing fraction = 0.58 \cite{Pusey2009Phil}) of the sediment; this can be used to calculate an effective particle radius for the silica ($R_{\mathrm{e}}$). If we assume that $R_{\mathrm{e}}$ is the sum of the average particle radius ($R$), Debye length ($\kappa^{-1}$) and standard deviation from polydispersity \cite{pusey1987effect}, the Debye length can be estimated to be $\approx$ 18 nm. Multiplying the inverse Debye length with the average radius of the particles, we find $\kappa R \approx 12 $. This value is very close to the $\kappa R$ for hard-sphere like suspensions \cite{yethiraj2003colloidal}, so that comparison of our results to those in hard-sphere systems are indeed useful here.

\section{Summary and Conclusions}
\par To summarize, we have captured and quantified the effect of increasing size polydispersity in thin films of colloidal crystals grown using vertical drying of silica particles with polydispersities of 6-15\% at narrowly spaced intervals. The samples have been quantitatively characterized here using electron microscopy combined with image analysis, FIB-SEM, and UV-Vis transmission spectra. The microscopic long-range and local structural order of the top layer, as well as the 3D order in interplanar spacing, decrease as the polydispersity of the assembling particles is increased. The 2D local six-fold bond-order parameter for the top layer shows a significant decline at 8\% and 12\% polydispersity. We have argued that the drop around 12\% can be explained by an argument analogous to the Lindemann criterion \cite{pusey1987effect}. We have provided two explanations for the drop around 8\%. First, we have observed little to no size fractionation, and it has previously been reported that crystallization requires fractionation above around 8\% polydispersity. Second, it has been reported that convective assembly can be considered as epitaxial growth of 2D crystalline layers, and 2D crystallization is inhibited above around 8\% polydispersity. Our observation that the local order is higher for crystalline layers further from the substrate provides support for the second explanation. Our results contribute to a systematic understanding of the formation of colloidal-crystal thin films from more polydisperse systems, for example synthesized through more sustainable methods.
\section{Author contributions}
MA: Writing original draft, investigation, data curation, formal analysis, visualization, writing - review \& editing. ABS: Investigation, resources, writing - review \& editing. FHJL: Investigation, resources, writing - review \& editing.  WCKP: Supervision, writing - review \& editing, funding acquisition.  JHJT: Conceptualization, methodology, validation, writing - review \& editing, supervision, project administration, funding acquisition. 
\section{Conflicts of interest}
There are no conflicts to declare.
\section{Acknowledgments}
M.A. acknowledges P. G. Iglesias for providing training on the use of the UV-Vis spectrophotometer and C. Stock for the use of his sputter coater to prepare assembly samples for error measurement. The electron microscope used in this project was funded by EPSRC CapEx investment through SOFI CDT (EP/L015536/1). We also acknowledge funding from EPSRC (EP/P030564/1) for the Cryo-FIB-SEM. For the purpose of open access, the author has applied a Creative Commons Attribution (CC BY) licence to any Author Accepted Manuscript version arising from this submission.

\clearpage

\appendix

\setcounter{figure}{0}
\renewcommand{\thefigure}{S\arabic{figure}}

\setcounter{table}{0}
\renewcommand{\thetable}{S\arabic{table}}

\makeatletter
\renewenvironment{widetext}{
   \begin{strip}
   \rule{\textwidth}{0pt}
}{
   \rule{\textwidth}{0pt}
   \end{strip}
}
\makeatother

\section*{Supplementary Information}

\subsection*{Stöber synthesis of silica nanoparticles}

\begin{table}[h!]
\centering
\begin{tabular}{|c|c|c|c|c|c|c|}
\hline
Batch & Ethanol & TEOS & Ammonia & Water \\
& (ml) & (ml) & (35\% in water, ml) & (distilled, ml) \\
\hline
1         & 1600     & 56 & 85 & 112 \\
2         & 1600     & 56 & 85 & 112 \\
3         & 1600     & 56 & 85 & 112 \\
5         & 1600     & 48 & 85 & 112 \\
7         & 1650     & 66 & 100 & 40 \\
8         & 1650     & 76 & 100 & 40 \\

\hline
\end{tabular}
\caption{Synthesis recipes of silica batches}
\label{tab:supplementary_table1}
\end{table}
Note: Batches 1-3 were prepared using the same procedure except TEOS was added gradually (over a few seconds) during the preparation of batch 3. The amount of TEOS was increased when going from batch 7 to 8, resulting in a slight increase in average particle size ($<$1 nm from  SEM measurement and 6 nm from DLS).
\begin{table}[h!]
\begin{tabular}{|c|l|}
\hline
Batch & \multicolumn{1}{c|}{Recipe} \\ 
\hline
4    &  Mixed batches 3 and 5 in a number concentration\\
     &  ratio of 1:1.14.\\
6    &  Mixed batches 2 and 3 in a number concentration\\
     &  ratio of 1:1.03.\\
\hline
\end{tabular}
\caption{Recipes of mixed silica batches}
\label{tab:supplementary_table2}
\end{table}
\vspace{-30pt}
\subsection*{ Moments of measured particle-size distributions
}

\begin{table}[H]
\centering
\begin{tabular}{|c|c|c|c|}
\hline
Batch & SEM size (nm) & Skew & Excess Kurtosis \\ \hline
1  & 384 & 0.006 & 1.68 \\ \hline
2  & 322 & -0.582 & 1.92 \\ \hline
3  & 381 & -0.838 & 4.31 \\ \hline
4  & 371 & -0.217 & -0.03 \\ \hline
5  & 365 & -0.128 &  0.25 \\ \hline
6  & 360 & 0.024 &  -0.36\\ \hline
7  & 208 & 0.235 & -0.19 \\ \hline
8  & 208 &  0.709& 1.03 \\ \hline
\end{tabular}
\caption{Average particle diameter, skew and excess kurtosis for particle-size distributions of the eight silica batches included in this study.}
\label{tab:supplementary_table3}
\end{table}

\subsection*{Estimation of surface potential of silica batches}
 \par To compare self-assembling behaviours, it is important to ensure that other intrinsic particle properties, such as surface potential remains consistent across batches. To compare to relevant literature, it is important to understand how `soft' the interparticle potential is. For practical purposes, zeta potential measurements are used as a proxy to estimate the surface potential of the batches. This approach has shown consistency with theoretical predictions and other experimental methods for determining silica surface potential \cite{horiuchi2012calculation}. Zeta potential for three silica batches (`low'=6.3\%, `intermediate'= 9.1\% and `high'= 14.6\% polydispersity) is measured by Electrophoretic Light Scattering (ELS) at 25\textdegree C using the Anton Paar Litesizer-500. An optically clear silica solution is formed by adding a droplet of dispersion (10-12\% silica in ethanol) to 15 ml ethanol and shaking well to ensure the particles are well dispersed. This solution is added to the measurement cell and then discarded to prepare the cell before refilling and taking measurements. The Smoluchowski approximation is used for the Henry function (f($\kappa$a)=1.5), which has been used as a valid approximation to determine zeta potential for metal oxides in organic solvents \cite{dembek2022solvent, kosmulski2020effects}. The zeta potentials for low, intermediate and high polydispersity silica batches are -47.5 mV, -51.2 mV and -48.3 mV. The colloid surfaces are negatively charged (as expected) and the dispersions are stable. The zeta potential values lie close to one another, implying that surface potential remains consistent across batches and is unlikely to influence differences in self-assembly behaviour.

\subsection*{Variation of average local six-fold bond-order parameter with average particle diameter}
\begin{figure}[H]
\centering
\includegraphics[width=0.48\textwidth]{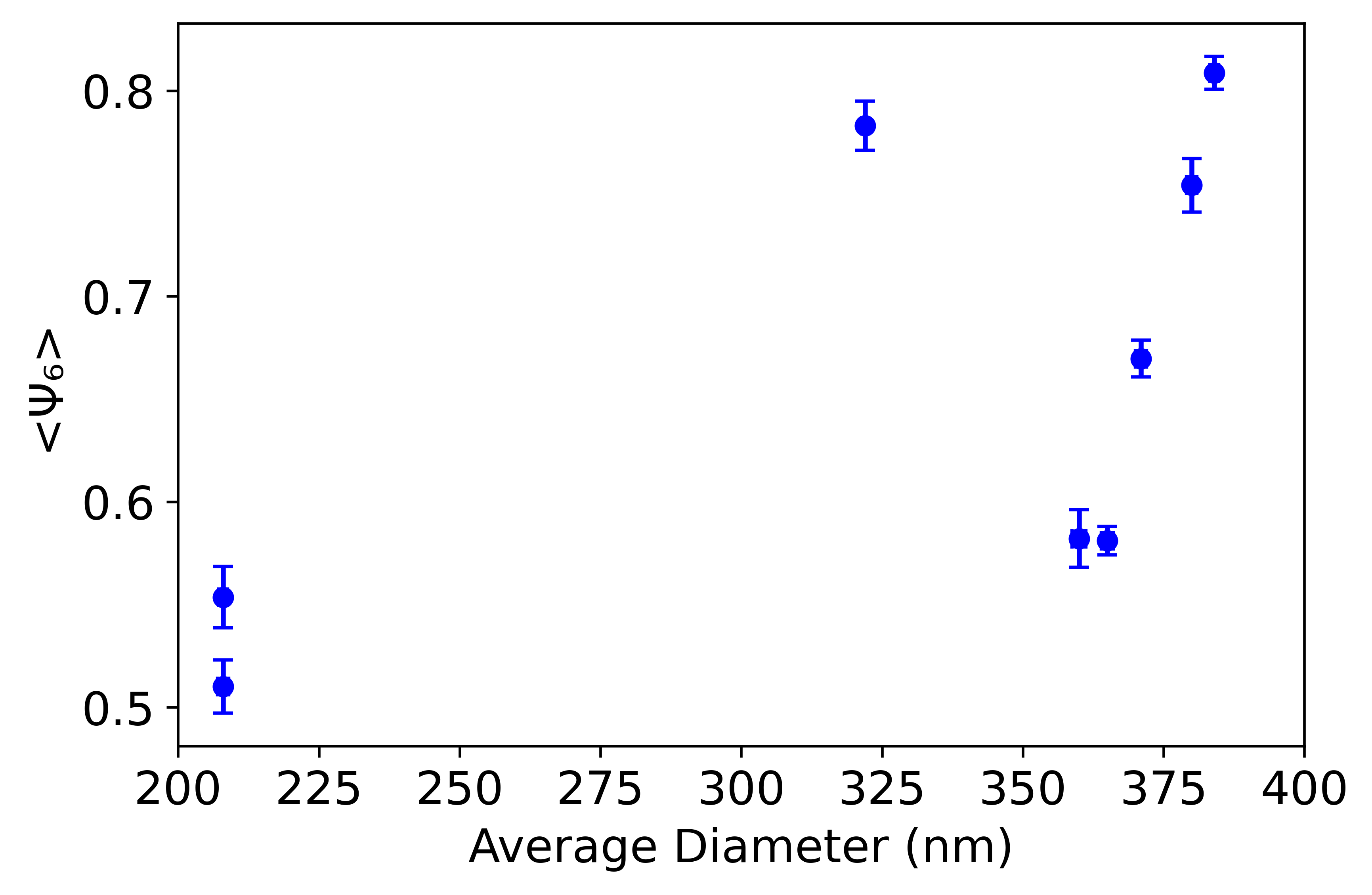}
\caption{Average particle diameter of silica batches vs. local six-fold bond-order parameter of their assemblies.}
\end{figure}
\end{document}